\documentclass[mathleft
]{an}
\usepackage{graphicx}
\usepackage{times}
\begin{document}

% The following seven commands are intended for editorial usage and should be ignored by
% the author(s).
\Pagespan{1}{}% Document's page range. 
% If second parameter is left empty, the last page is computed automatically.
\Yearpublication{2008}%
\Yearsubmission{2008}%
\Month{}%   
\Volume{}%  
\Issue{}% 
\DOI{}
%{This.is/not.aDOI}% 

\title{Status and Perspectives of Astroparticle Physics in Europe}

\author{Christian Spiering 
%\inst{1}\fnmsep
\thanks{\email{christian.spiering@desy.de}
\newline}
%Example 
%for footnote, note the usage of the \texttt{fnmsep}
%command as separator between institute number and footnote mark} 
}
%\titlerunning{Instructions for authors}
%\authorrunning{T.H.E. Editor \& G.H. Ostwriter}
\institute{
DESY, Platanenallee 6, D-15738 Zeuthen, Germany}

\received{February 2008}
%\accepted{2008}
\publonline{2008}


\abstract{
Astroparticle physics has evolved as an interdisciplinary field at the 
intersection of particle physics, astronomy and cosmology. Over the last 
two decades, it has moved from infancy to technological maturity and is 
now envisaging projects on the 100 Million Euro scale. This price tag requires 
international coordination, cooperation and convergence to a few flagship 
projects. The Roadmap Committee of ApPEC (Astroparticle Physics European 
Coordination) has recently released a roadmap covering the next ten years. 
This talk describes status and perspectives of astroparticle physics in 
Europe and reports the recommendations of the Roadmap Committee.
} 

\maketitle

\section{Introduction}

Although the notation "astroparticle physics" was coined only 
25 years ago, and although it has been widely used only since the nineties, 
its roots go back to the early years of the last century, when Victor Hess 
discovered cosmic rays. The origin of these particles -- mostly protons, 
light and heavy nuclei -- was unknown, and it is going to be solved only now, 
100 years later. However, there was a long period when cosmic rays served 
as the source of most newly discovered elementary particles: 1932, the first 
anti-particle, the positron, was recorded in cosmic rays, followed by the 
muon in 1936 and in 1947 by the pion, the first of the vast family of mesons. 
Until the beginning of the fifties, cosmic rays remained the main source of 
new particles and laid the ground for the "'particle zoo"', which a decade 
later was explained by the quark model. It was only in the mid of the fifties 
that particle accelerators started their triumphal development and cosmic rays 
lost their role in particle physics. With a single 
exception, cosmic particle physics disappeared from the screen of most 
particle physicists. 

The exception was the long-lasting attempt to detect solar neutrinos. 
It was pioneered by Ray Davis in the sixties
who measured the $^8$B neutrino flux
with the radio-chemical ClAr method -- detecting however  
only a third of the predicted flux. 
The deficit was confirmed  in the eighties by Kamiokande, a water
Cherenkov detector in Japan. The following measurements of pp neutrinos
with GaGe detectors in the Russian Baksan laboratory (SAGE experiment) and
in the Italian Gran Sasso Laboratory (GALLEX experiment)
corroborated the suggestion that the solution to the neutrino
deficit was not given by a different solar model but by neutrino oscillations.
This solution was eventually confirmed by the SNO experiment in Canada 
(Bahcall 2005, McDonald et al. 2004).

In 2002, Ray Davis and Masatoshi Koshiba were 
awar\-ded the Nobel Prize in Physics 
for opening the neutrino window to the Universe, specifically for the 
detection of neutrinos from the Sun and the Supernova SN1987A in the 
Large Magellanic Cloud. Their work was a unique synthesis of particle physics 
and astrophysics since solar neutrinos also provided the first clear evidence 
that neutrinos have mass. It
represents a classical illustration of the 
interdisciplinary field at the intersection of particle physics, astronomy 
and cosmology which now is known as astroparticle physics.
One may note that Koshiba's Kamiokande detector was originally built to detect 
proton decay -- another bracketing of astrophysics and particle physics by a 
single technique.

The detection of solar and Supernova neutrinos is not the only new window 
to the Universe opened by astroparticle physics. Another one is that of high 
energetic gamma rays recorded by ground based Cherenkov telescopes. 
From the first source detected in 1989, three sources known in 1996, to nearly 
70 sources identified by the end of 2007, the high energy sky has revealed a 
stunning richness of new phenomena and 
puzzling details (see Fig.~\ref{gammasky} below). 
Other branches of astroparticle physics did not yet provide such gold-plated 
discoveries but moved into unprecedented sensitivity regions with rapidly 
increasing discovery potential, like the search for dark matter particles, 
the search for decaying protons or the attempt to determine the absolute 
values of neutrino masses.

\section{Basic questions}

The Roadmap Committee of ApPEC (Astroparticle Physics European Coordination) 
has recently released a roadmap covering the next ten years. Recommendations 
of the committee (http://www.aspera-eu.org) have been formulated by addressing 
a set of basic questions:

\begin{enumerate}

\item What are the constituents of the Universe? 
In particular: What is dark matter?
\item Do protons have a finite life time?
\item What are the properties of neutrinos? 
What is their role in cosmic evolution?
\item What do neutrinos tell us about the interior of the Sun and the 
Earth, and about  
       Supernova explosions?
\item What is the origin of cosmic rays ? What is the view of the sky at 
extreme energies ?
\item  What will gravitational waves they tell us about violent cosmic 
 processes and about the nature of gravity?

\end{enumerate}

An answer to any of these questions would mark a major 
break-through in understanding 
the Universe and would open an entirely new field of research on its own.

\section{Dark Matter and Dark Energy}

Over the last decade, the content of the Universe has been 
measured with unprecedented precision. 
Whereas normal baryonic matter contributes only about 4\%, 
the dominant constituents are unknown forms of matter and energy: 
Dark Matter (22\%) and Dark Energy (74\%).

Whereas the concept of Dark Energy was introduced only recently 
-- in response to a negative pressure driving cosmic expansion --, 
Dark Matter has been discussed for decades. 
The prevalent view is that Dark Matter consists of stable relic 
particles from the Big Bang, and that nearly all of it is in the 
form of Cold Dark Matter (CDM). In the early Universe, CDM particles 
would have already cooled to non-relativistic velocities 
when decoupling from the expanding and cooling Universe. 
Hot dark matter (HDM) has been relativistic at the time of decoupling. 
Neutrinos are typical HDM particles; their contribution to the total 
matter budget, however, is small.

\subsection{The Search for Dark Matter}             
                  
The favoured candidate for dark matter is a Weakly Interacting Massive 
Particle (WIMP) related to new physics at the TeV scale (Steigmann \& Turner
1985, Jungmann et al., 1996). Among the 
various WIMP candidates, the 
lightest supersymmetric (SUSY) particle in
the Minimal SuperSymmetric Model (MSSM) is favoured -- likely
the {\it neutralino}.  
Another theoretically well-founded dark matter candidate is the axion
(Peccei \& Quinn 1977, Raffelt 2006). 
Even though axions would be much lighter than WIMPs, they still 
could constitute CDM, since they are have not been produced in 
thermal equilibrium and would be non-relativistic.

\vspace{3mm}
\noindent
{\it Direct WIMP searches}
\vspace{1mm}

\begin{figure}[h]
\includegraphics[width=.45\textwidth]{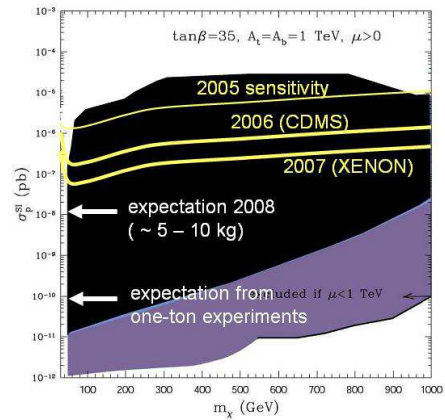}
\caption{
%\label{darkmatter}
Spin-independent WIMP cross section vs. WIMP mass for an MSSM prediction 
(dark area), with parameters 
fixed to the values shown at top right (Kim et al. 2002). 
The upper curve represents the 
limits obtained by 2005. CDMS, a bolometric detector operated in the US, 
achieved the 2006 record limit and was surpassed in 2007 by XENON, a 
liquid xenon detector operated in the Italian Gran Sasso Underground 
Laboratory (Angle 2008).  
The arrows indicate the sensitivities expected in about a year 
from now, and within a decade from now with one-ton experiments. All results 
assume the WIMP to be a neutralino in the standard MSSM formulation.
}
\label{darkmatter}
\end{figure}

"Direct" WIMP searches focus on the 
detection of nuclear recoils 
from WIMPs interacting in underground detectors (Gaitskell 2004,
Baudis 2006, Sadoulet 2007). 
No WIMP candidate has been found so far. Assuming that all Dark Matter is 
made of WIMPs, present experiments with a several-kilogram 
target mass can therefore exclude 
WIMPS with an interaction cross section larger than $10^{-43}$ cm$^2$ 
(i.e. $10^{-7}$ picobarn). 
MSSM predictions for neutralino cross sections range from  
$10^{-5}$ to $10^{-12}$ pb (see Fig.\ref{darkmatter}). 

Experimental sensitivities will be boosted to $10^{-8}$ pb
in a couple of years and 
may reach, with ton-scale detectors, $10^{-10}$ pb
in 7-10 years. Therefore, 
there is a fair chance to detect dark matter particles in the next decade -- 
provided the progress in background rejection can be realized and provided 
CDM is made of super-symmetric particles. 

Presently, there are two favoured detection techniques:
  
{\it Bolometric} detectors 
are operated at a temperature of 10-20 mK and detect the 
feeble heat, ionization 
and scintillation signals from WIMP interactions in crystals made, e.g., 
from germanium, silicon or CaWO$_4$. Present flagship experiments are
CDMS in the USA, and CRESST (Gran Sasso Laboratory, Italy) 
and EDELWEISS (Fr\'ejus Laboratory, France) in Europe.

{\it Noble liquid} detectors record ionization 
and scintillation from nuclear recoils in liquid xenon, argon or neon. 
XENON (Gran Sasso) and ZEPLIN (Boulby mine, UK) use liquid xenon
targets of about 10kg mass, while
WARP (Gran Sasso) and ArDM (Canfranc, Spain) operate, or prepare,
liquid argon detectors.  Actually the most recent significant step in 
the race for better sensitivities 
has been made by XENON (see Angle 2008 and Fig.\ref{darkmatter}).

\begin{figure}[h]
\includegraphics[width=.45\textwidth]{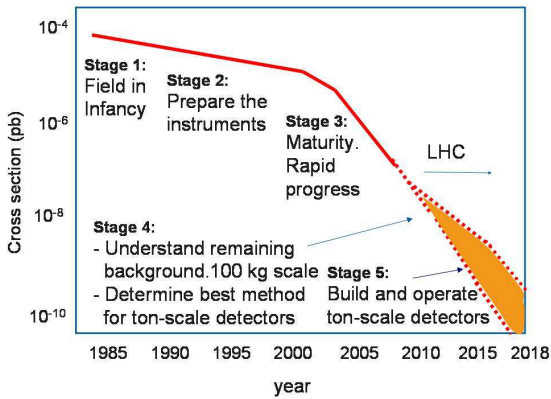}
\caption{
%\label{DM-Trend}
Possible development of limits and sensitivities as a function
of time (see text for explanations).
}
\label{DM-Trend}
\end{figure}

\begin{figure}[h]
\includegraphics[width=.42\textwidth]{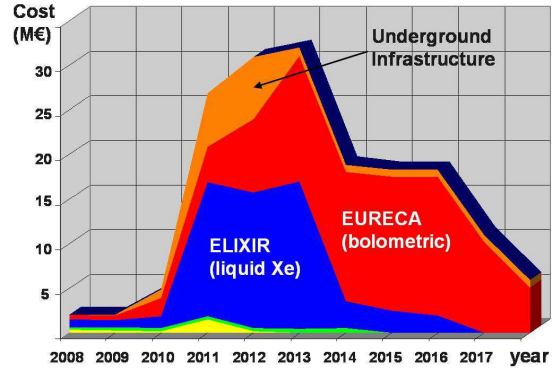}
\caption{
%\label{DM-cost}
Present projection of the investment expenses for the ton scale versions
of ELIXIR and EURECA, including the cost for an low-background underground
infrastructure housing the detectors. Since detectors will be built
in a modular way, this scenario includes a stepwise approach to the 
ton scale, via 100-kg stages. 
}
\label{DM-cost}
\end{figure}

A variety of presently more than 20 dark matter experiments worldwide 
(see for a review Baudis 2007) must, 
within several years, converge to two or three few ton-scale experiments with 
negligible background. In Europe, there 
are ~two large initiatives ~towards 
~experiments ~on the ton scale: EURECA, 
joining most players of the bolometric 
approach, and ELIXIR, joining most of the liquid xenon experts. R\&D on 
alternative methods will be continued, but most of the resources 
will naturally 
be focused to ton-scale flagship pro\-jects and the corresponding underground 
infrastructures.
Figures \ref{DM-Trend} and \ref{DM-cost} sketch a scenario towards
ton-scale experiments with negligible background (different to ton-scale
experiments attempting to identify an annual signal variation
on top of a large background, like the DAMA experiment, see below). 
Figure \ref{DM-Trend}
shows the possible development of limits and sensitivities as a function
of time, assuming a standard MSSM WIMP with spin-independent coupling.
A $10^{-8}$\,pb sensitivity can be reached within the next couple of years.
Improvements by further two orders of magnitude require 
more massive detectors. The coloured area
for $>$2009 indicates the range of projections given by
different experiments, most of them envisaging an intermediate step
at the 100 kg scale. Note that this scenario is made from a 2007 perspective
and that initial LHC results may substantially influence the design
of the very few "ultimate" detectors. Needless to say that all these
plans stand or fall with the capability to reduce the background,
even for ton scale masses, to less than a very few events per year.

Figure \ref{DM-cost} shows a first tentative
projection of investment expenses for
ELIXIR and EURECA, including the cost for construction ~of a
~suitable ~low-background, ~deep under\-ground infrastructure 
\footnote{This figure may serve as an illustration for
the kind of information which has been collected for all astroparticle
experiments in Europe.}.
Estimates will be made more precise within design phases
~covering the ~next three years. ~Note that a possible
priorisation within given
funding envelopes may change this picture substantially.

Following identification of a proper signal, clearly distinguished against 
background, one would like to get a final confirmation for the nature of 
the signal, by observing a "'smoking gun"' signature which ensures that the 
signal is due to WIMPs and not due to something else, such as backgrounds. 
There are three such signatures: 
{\it a)} annual modulation, {\it b)} directionality, 
{\it c)} target dependence.

\begin{figure}[h]
\center{
\includegraphics[height=.23\textwidth,width=.45\textwidth]{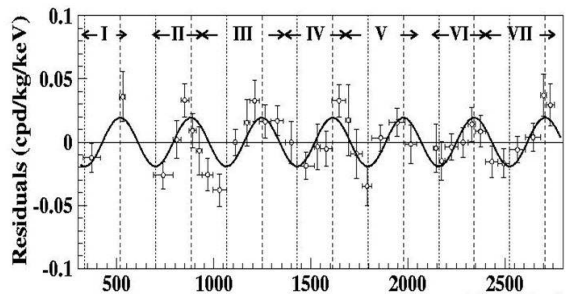}
}
\caption{
%\label{dama}
Seasonal variations of the counting rate residuals observed
by the DAMA experiment over a period of seven years (Bernabei 2004).
}
\label{dama}
\end{figure}

The annual modulation signature reflects the periodic change of the WIMP 
velocity in the detector frame due to the motion of the Earth around the Sun. 
The variation is only of a few percent of the total WIMP signal, therefore 
large target masses are needed to be sensitive to the effect. Indeed, the 
DAMA experiment (Bernabei 2004), 
recording the  scintillation light in NaI 
crystals, has reported an observation of this signature in its data
from 100 kg NaI (see Fig.\ref{dama}), but the 
interpretation remains controversial. The collaboration is
presently running a 250 kg version of the experiment, with
first results expected in 2008, and is asking for ton-scale
ressources in a next step.

The target dependence signature 
follows from the different interactions of WIMPs with different nuclei - both 
in rate and in spectral shape. 
The directionality signature would clearly 
distinguish the WIMP signal from a terrestrial background and search for a 
large forward/backward asymmetry. It requires detectors capable of measuring 
the nuclear recoil direction, a condition potentially only met by 
gaseous detectors. At present, these detectors are in an R\&D phase.

\vspace{3mm}
\noindent
{\it Indirect Dark Matter Search}
\vspace{1mm}

The mentioned {\it direct} searches are flanked by {\it indirect searches}. 
These would identify charged particles, gamma rays or neutrinos from WIMP 
annihilation in the cores of galaxies, the Sun or the Earth. WIMPs can be 
gravitationally trapped in celestial bodies and eventually accumulate until 
their density becomes large enough that the self-annihilation of WIMPs would 
be in equilibrium with WIMP capture. Earth-bound or satellite detectors would 
then detect the decay products of these annihilations: an excess of neutrinos 
from the centre of the Earth or Sun, a gamma signal from the centre of the 
Galaxy, or an excess in positrons and anti-protons from galactic plane or 
halo. 
Anti-particles such as positrons and anti-protons produced in WIMP annihilation 
would be trapped in the galactic magnetic fields and be detected as an excess 
over the background generated by other well understood processes.

Present and planned detectors capable to contribute to indirect
Dark Matter detection are described in section 7: satellite detectors
(Pamela and AMS for charged cosmic rays, Agile and GLAST for gamma rays)
and earth bound telescopes (Magic, H.E.S.S., Veritas, CTA for gamma rays,
Baikal, IceCube, Antares, KM3NeT for neutrinos).

Direct and indirect methods give complementary information. For instance, 
gravitational trapping works best for slow WIMPs, making indirect searches 
most  sensitive. In the case of direct detection, 
however, the higher energy of recoil 
nuclei makes fast WIMPs easier to detect. 
For indirect searches, the annihilation 
rates would depend on all the cosmic history of WIMP accumulation and not only 
on the present density, providing another aspect of complementarity to 
direct searches. 

\vspace{3mm}
\noindent
{\it Synthesis of direct, indirect and accelerator signatures}
\vspace{1mm}

While astroparticle physicists are searching dark matter, particle physicists 
are preparing searches for super-symmet\-ric particles at  
the Large Hadron Collider in Geneva, which is 
expected to start operation end of 2008 and may provide
first physics results on SUSY searches in 2010 or 2011.  
A detection of SUSY
particles at the LHC would certainly considerably boost
dark matter searches. Eventually,
only the synthesis of all three observations --
direct and indirect detection of cosmic candidates for 
dark matter and identification 
of the neutralino at the LHC -- would give sufficient 
confidence about the character 
of the observed particles. 
(see the recent review of Bertone 2007 for a description
of a multidisciplinary approach to Dark Matter search).  

\subsection{Dark Energy}

Evidence of Dark Matter and Dark Energy has emerged from astronomical
observations. ~~Astronomical studies of galactic dynamics, gravitational lensing,
large scale structures and CMBR anisotropies provide arguments for
Dark Matter. Combining these observations with the observation
that the universe is accelerating (SNIa methods), that the Universe is
flat (from CMBR measurements) and that Dark Matter alone cannot provide 
the critical density (from large scale structures) ~establishes the
need ~for something like "Dark Energy". 

However, whereas Dark Matter may consist of distinct
particles and can be searched by the 
methods of astroparticle
physics, Dark Energy may be a continuous phenomenon.
``Particle search strategies'' equivalent to the Dark
Matter case do not exist.
~~Dark Energy can primarily explored through its influence on
cosmic evolution. Observations in this area traditionally use
astronomical techniques (see e.g. Perlmutter \& Schmidt 2003, 
Spergel et al. 2007). 
Particle physicists have joined
this new field and are playing a major role -- for instance by
contributing with their experience in processing large amounts of data.

The next generation 
of experiments relevant for Dark Energy search 
includes the European Planck mission on a satellite, and 
the ground based Dark Energy Survey, DES, the
Low Frequency Array, LOFAR, ~and the ALMA-Path\-finder APEX. 
Projects proposed to be started after 2013 include various
survey telescopes, most notably
the Large Synoptic Survey Telescope, LSST, and the
Panoramic Survey Telescope \& Rapid Response System,
~PanSTARRS. ~~Space based missions include the
the wide field space imager DUNE and the
spectroscopic all-sky cosmic explorer, SPACE,
the
SuperNova/Acceleration Probe, SNAP, and
the James Webb Space Telescope, JWST.
Needless to say that 
survey telescopes serve a variety of standard astronomical
tasks and are not bounded to Dark Energy search. This
is also true for the European Extremely Large
Telescope, E-ELT, and the Square Kilometer Array, SKA.
The inclusion of SKA in the ESFRI list demonstrates
a high European priority. 

Given the deep implications for fundamental physics,
Dark Energy missions find the strongest support from
the astroparticle physics community. ~~Recommendations
are formulated 
in the European ASTRONET Roadmap.

\section{Proton decay and low energy neutrino astronomy}

Grand Unified Theories (GUTs) of particle physics predict that 
the proton has a 
finite lifetime. The related physics may be closely linked to 
the physics of the 
Big Bang and the cosmic matter-antimatter asymmetry. 
Data from the Super-Kamiokande 
detector in Japan constrain the proton lifetime to be 
larger than 10$^{34}$ years, 
tantalizingly close to predictions of various SUSY-GUT predictions. 
A sensitivity improvement of an order of magnitude requires detectors on the 
10$^5$-10$^6$ ton scale. 
The discovery of proton decay would be one of the most 
fundamental discoveries for physics and cosmology and certainly merits a 
worldwide coherent effort.
 
Proton decay detectors do also detect cosmic neutrinos. 
Figure 5 
%\ref{GuNu} 
shows a "grand 
unified neutrino spectrum". Solar neutrinos, burst neutrinos from SN1987A, 
reactor neutrinos, terrestrial neutrinos and atmospheric neutrinos have been 
already detected. They would be also in the focus of a next-stage proton 
decay detector. Another guaranteed, although not yet detected, flux is that 
of neutrinos generated in collisions of ultra-energetic protons with the 3K 
cosmic microwave background (CMB), the so-called GZK (Greisen-Zatsepin-Kuzmin) 
neutrinos. ~~Whereas GZK neutrinos as well as 
neutrinos from Active Galactic Nuclei 
(AGN) are likely to be detected by neutrino telescopes in the coming decade 
(see below), no realistic 
idea exists how to detect 1.9 K cosmological neutrinos, 
the analogue to the 2.7 K microwave radiation.

\begin{figure}[h]
\center{
\includegraphics[width=.45\textwidth]{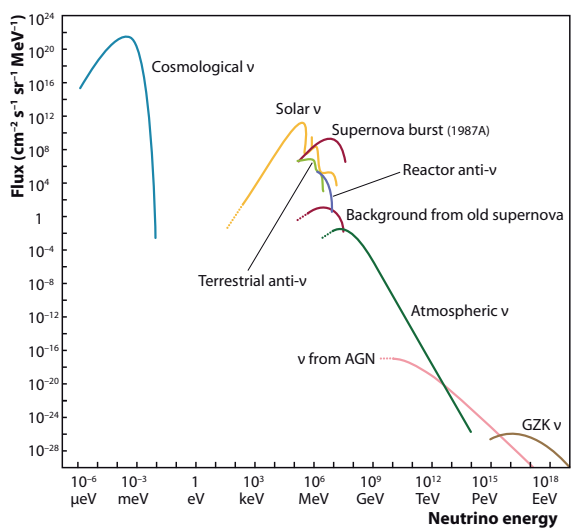}
\caption{The "grand unified" neutrino spectrum}
}
\label{GuNu}
\end{figure}

Solar neutrinos, detection of neutrino oscillations by solar 
and atmospheric neutrinos, 
neutrinos from the supernova SN1987A, geo-neutrinos -- large 
underground detectors have 
produced an extremely rich harvest of discoveries (McDonald et al. 2004). 
We note the most recent event in the series of solar neutrino
measurements: the first real-time detection of solar $^7$Be neutrinos
by the BOREXINO experiment (Arpesella et al. 2008)
-- an impressive success, 
eventually achieved after a long troublesome process
of internal, and in particular externally imposed, delays. 

The triumphal legacy of underground neutrino physics is 
intended to be continued by worldwide one or two multi-purpose detectors on 
the mass scale of 100-1000 kilotons. The physics potential of such a large 
multi-purpose facility would cover a large variety of questions:

\begin{enumerate}

\item The proton decay sensitivity would be improved by one order of magnitude.
\item A galactic Supernova would result in 10$^4$-10$^5$ neutrino 
events, compared to 
only 20 events for SN1987A. 
This would provide incredibly detailed information on 
the early phase of the Supernova explosion.
\item The diffuse flux from past supernovae 
would probe the cosmological star formation rate.
\item The details of the processes in the solar interior 
can be studied with high 
statistics and the details of the Standard Solar Model 
determined with percent accuracy.
\item The high-statistics study of atmospheric neutrinos could improve our 
knowledge on the neutrino mass matrix and provide unique information on the 
neutrino mass hierarchy.
\item Our understanding of the Earth interior would be improved by the 
study of geo-neutrinos.
\item The study of neutrinos of medium energy from the 
Sun and the centre of the Earth 
could reveal signs for dark matter.
\item Last but not least, a large underground detector could detect 
artificially 
produced neutrinos from nuclear reactors or particle accelerators, over a long 
baseline between neutrino source and detector. 
\end{enumerate}

Three detection techniques are currently studied: Water-Cherenkov detectors 
(like Super-Kamiokande), liquid scintillator 
detectors (like BOREXINO) and liquid argon detectors (a technique 
pioneered by the Italian ICARUS collaboration). 
The present prominent European projects under study are 
MEMPHYS, a Megaton-scale water detector (de Bellefon et al. 2006), 
LENA (Wurm et al. 2007), a 30-70 kiloton liquid scintillator detector, and
GLACIER (Rubbia 2004), a 50-100 kiloton liquid argon detector
(see Fig.6).
%\ref{threetechniques}).

\begin{figure}[h]
\center{
\includegraphics[width=.45\textwidth]{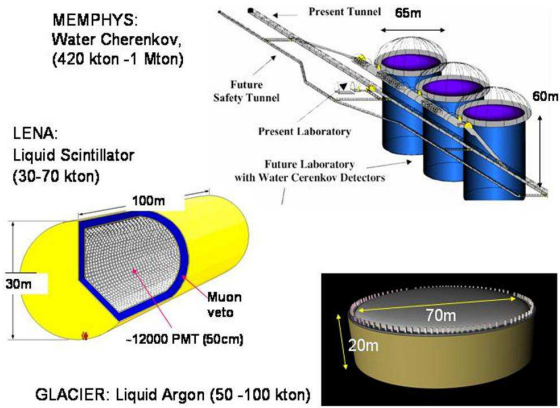}
\caption{Representative projects for the three proposed methods: 
MEMPHYS (420-1000 kton water), LENA (30-70 kton scintillator)  
and GLACIER (50-100 kton liquid argon).}
}
\label{threetechniques}
\end{figure}

\noindent
The European LAGUNA consortium (Autiero et al. 2007)
has been awarded a FP7 Design Study grant in order 
{\it a)} to explore and compare 
the capabilities of the three methods and {\it b)} 
to evaluate the possibilities 
of excavation of deep large cavities and the accompanying infrastructures. 
The design study should converge, on a time scale of 2010, 
to a common proposal.  
The total cost depends on the method and the actual size and 
is estimated to be between 
400 and 800 MEuro. 
Civil engineering may start in 2013. The cost
would be shared internationally for such a Mega project.

\section{Neutrino properties}

In the context of astroparticle physics, neutrinos, 
rather than being the subject of 
research, mainly play the role of messengers: 
from the Sun, from a Supernova, from 
Active Gala\-xies and other celestial objects. 
Still, some of their intrinsic properties remain undetermined. From the 
oscillatory behaviour of neutrinos we can deduce that the weak eigenstates of 
neutrinos ~(flavour ~eigenstates) 
are not identical with their mass eigenstates, 
that the neutrino masses are non-zero and that they differ from each other. 
That is why the 
flavour eigenstates, electron neutrino $\nu_e$, muon neutrino $\nu_{\mu}$ 
and tau neutrino $\nu_{\tau}$,  
oscillate between each other. 
From oscillations we can determine how strong
the states mix and the mass differences. 
But what are the absolute values of the masses? 
Further: are neutrinos their own 
antiparticles ("Majorana particles")?  

Many of the projects devoted to these questions are 
not really {\it astro}-particle 
experiments, but often share certain aspects with ``typical'' astroparticle 
experiments: 
the infrastructure (like low-background, deep caverns), the methods 
(like high purity liquid scintillator techniques), or sometimes just the 
scientist's community. That is why they are addressed by the ApPEC roadmap. 

The main sources of information on neutrino parameters are the following:  

\begin{itemize}
\item[a)] oscillation experiments 
using neutrinos from accelerators or nuclear 
reactors as well as atmospheric or solar neutrinos. They provide information 
on mixing parameters, on a possible CP violation and on mass differences, 
but not absolute masses. 
\end{itemize}

Absolute masses can be derived from three types of data or experiments: 

\begin{itemize}
\item[b)] cosmological data (see for a review Turner 2007), 
\item[c)] end-point measurement of electron energy spectra in $\beta$-decays, 
\item[d)] neutrino-less double beta decay experiments. 
They are 
the only experiments which can also prove the Majorana nature of neutrinos.
\end{itemize}

Neutrino oscillations impose a lower limit on the 
heaviest neutrino mass of about 
0.05\,eV (since this occurs to be the mass difference between the heaviest and 
the second heaviest mass state). 
This implies that neutrinos contribute at least 
0.1\% of cosmic matter. 
Neutrinos with a small finite mass contribute to hot dark matter, 
which suppresses the power spectrum of density fluctuations 
in the early Universe 
at "'small"' scales, of the order of one to ten Mega-parsec. 
The recent high precision measurements of density fluctuations in the Cosmic 
Microwave Background (WMAP) and the observations of the Large Scale Structure 
distribution of galaxies (2dFGRS and SDSS), 
combined with other cosmological data, 
yield an upper limit of about 1.5\% on the amount of 
hot dark matter in the Universe, 
corresponding to an upper limit of about 0.6-0.7\,eV on 
the sum of all three neutrino 
masses (see e.g. Hannestad \& Raffelt 2006). 
The future sensitivity of cosmological measurements with Large Scale 
Surveys and with the CMB mission Planck, combined with the weak gravitational 
lensing of radiation from background galaxies 
and of the CMB is expected to reach 
a value of  $\approx$\,0.1\,eV. 

\vspace{3mm}
\noindent
{\it Direct mass measurement}
\vspace{1mm}

The only laboratory technique for the direct measurement of a small neutrino 
mass (without additional assumptions on the character of the neutrino) is the 
precise measurement of the electron spectrum in $\beta$-decays. 
Here, the neutrino mass 
(or an upper limit to it) is 
inferred from the shape of the energy spectrum near 
its kinematical end point. The present upper limit is at 2.3\,eV
(Kraus et al., 2005, Lobashov et al. 2003). The KATRIN 
experiment in Karlsruhe will improve the 
sensitivity of past experiments down to 
0.2\,eV (Robertson 2007). 
Operation of KATRIN is expected to start in 2009/2010. 

Given the cosmological sensitivities, 
one may ask for the competitiveness of the KATRIN experiment.  
Precision cosmology yields an upper limit of 0.7\,eV for the sum of all three 
neutrino masses (or 0.7\,eV/3\,$\approx$\,0.23\,eV with respect 
to the lightest mass state). 
This does not seem to leave much room for a device with 0.2-eV sensitivity. 
However, one must keep in mind that the cosmological limit, 
despite the impressive 
success of precision cosmology, has to be derived within 
a system of assumptions 
and interpretations, and is not obtained directly. Considering the importance 
of the neutrino mass question, 
and the difficulty in associating the cosmological 
limit to a precise systematic confidence level, 
it is therefore important to pursue 
direct measurements up to their eventual 
technological -- and financial -- limits. There
is only one way to move beyond KATRIN sensitivity:
using calorimetric instead of spectrometric methods.
The potential of these
methods is presently explored.

\vspace{3mm}
\noindent
{\it Neutrino-less double beta decay}
\vspace{1mm}

The observation of neutrino-less double beta decay 
may allow going to even lower 
masses than end-point measurements of the KATRIN type. 
However, it requires the 
neutrino to be a {\it Majorana} particle, 
i.e. representing the only fermion which is 
its own anti-particle. Implications of massive neutrinos for models beyond the 
Standard Model differ for Majorana and Dirac neutrinos. Therefore the answer 
to the question whether nature took the "Majorana option" is essential. 

In a neutrino-less double 
beta decay, a nucleus $(A,Z)$ would turn into another 
$(A,Z+2)$ by transforming two neutrons into protons and 
emitting two electrons: 
$(A,Z) \rightarrow (A,Z+2) + 2e^-$ . 
This differs from "normal" double-beta decay 
(second order process of the weak interaction), which is rare but has been 
detected and studied: 
$(A,Z) \rightarrow (A,Z+2) + 2e^- + 2\nu$. Neutrino-less double 
beta decay is possible only for massive Majorana neutrinos. 
The observed lifetime would be inversely proportional 
to the neutrino mass squared. 
Corresponding experiments are performed in low radioactivity environments deep 
underground, in order to suppress fake events (see for overviews
Elliot \& Vogel 2002 and Avignone, Elliot \& Engel 2007).

Searches for double beta have been performed since the 1950s, 
~~but it was the 
discovery of ~neutrino 
oscillations which eventually led to a renaissance of the 
early enthusiasm and enormously boosted the existing efforts. Present best 
limits are at 0.3-0.8\,eV (Avignone,Elliot\&Engel\,2007), 
with the uncertainty in the mass 
limit reflecting the 
limited knowledge on the nuclear matrix elements. 
Figure 7
%\ref{DBD} 
shows the allowed effective neutrino masses (i.e.
the linear
combination of masses of the three mass states 
which is measured in double beta
experiments) vs. the mass of the lightest neutrino.
Constrains come from the mentioned double-beta limit and from
cosmological observations. 

\begin{figure}[h]
\center{
\includegraphics[width=.45\textwidth]{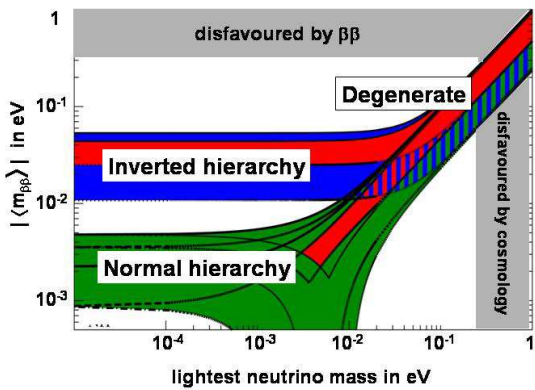}
\caption{
Allowed effective neutrino mass (as measured in double beta decay
experiments) vs. mass of the lightest mass state (adopted from Petcov 2005).
Different lines and colours for theoretical predictions correspond to
various assumptions on CP violating phases. Degenerate scenarios correspond
to nearly equal neutrino masses, for different masses one distinguishes
``normal'' hierarchies and ``inverted'' hierarchies}
}
\label{DBD}
\end{figure}

One single experiment ~(Klapdor-Kleingrothaus et al. 2004) 
claims a positive observation and derives a mass 
of 0.2-0.6 eV. This claim is highly controversial 
but can be tested with next 
generation experiments which will reach a 
sensitivity of better than 0.1 eV between 
2009 and 2013. At present
there are three such
European flagship projects: CUORICINO, a bolometric
detector in the Gran Sasso Lab,
uses 41 kg $^{130}$Te and
plans to operate a 740 kg detector (CUORE) in the next decade. NEMO
is a detector with both tracking and calometry capabilities using
8 kg  $^{130}$Mo and $^{62}$Se in the Fr\'ejus underground laboratory. 
A 100 kg $^{62}$Se detector (Super-NEMO) for the next decade
is in the design phase. 
GERDA is a $^{76}$Ge
detector.
%, i.e using the same nucleus like Klapdor-Kleingrothaus.
It will be operated in the Gran Sasso Lab from 2009 on
and stepwise upgraded from 15 kg to 35 kg. In another
approach, called COBRA, the use of CdTe is tested.
US projects include MAJORANA, a germanium project like GERDA,
and EXO, a xenon detector. These experiments could 
possibly ``scrape''
the mass range of the inverted hierarchy scenarios
in Fig.7.
%\ref{DBD}.

Europe is currently clearly
leading the field of double beta decay searches and is
in the strategic position to play a major role in next
generation experiments.
To cover a mass range of 
20-50\,meV, i.e. most of the range
suggested by the inverted mass scale scenarios, 
one needs detectors with an active mass of order one ton, 
good resolution and very low background. Construction of such 
detectors might 
start in 2014-2017. Different nuclear isotopes and different 
experimental techniques are needed to 
establish the effect and extract a neutrino 
mass value. 
The price tag for one of these experiments is 
at the 100-200 Million Euro scale, 
with a large contribution from the 
production cost for isotopes. The priority and urgency
with which these experiments will be tackled will 
depend on the background rejection achieved in the
currently prepared stages, on the available funding,
and on the future bounds on the neutrino mass from cosmological
observations.

\section{Underground Laboratories}

Proton decay, neutrino-less double beta decay or interaction
of dark matter particles are extremely rare processes and the effects
are extremely feeble. The signals from solar or geo-neutrinos
are similarly weak. The study of these processes
requires a low-background environment, shielded against
processes which may fake a true signal.
This environment is provided by special underground laboratories.
The various tasks require different characteristics of the site.
Double beta experiments and dark matter searches
need housing for ton-scale detectors and very
low radioactive background. Detectors for solar neutrino
neutrinos need a moderate depth between 1000 and 2000 meter water
equivalent and much larger caverns. For proton decay experiments,
neither large depth nor extremely low radiactivity is needed,
but the cavern for a Megaton detector naturally has to be huge.
  
There are five European underground laboratories which have been used in 
the past and are used presently for astroparticle physics deep underground 
experiments, ~~with depths ranging between one and nearly five kilometer 
water equivalent : The Laboratori Nazionali del Gran Sasso (LNGS) 
along a motorway tunnel in the Apennines (Italy), the Laboratoire 
Souterrain de Modane, LSM, ~located along the Fr\'ejus Road tunnel 
connecting Italy and France, ~the Laboratorio subterr\'aneo de Canfranc, LSC, 
arranged along a tunnel connecting 
Spain and France, the Boulby Underground 
Laboratory in an operational potash and 
rock-salt mine on the North-East coast of  
England.  Russia operates the Baksan Neutrino Observatory, BNO, 
in a dedicated tunnel in the Caucasus. 
A sixth very deep site in Finland is under discussion and some 
additional shallow locations are considered for special applications 
or as test sites (see Coccia 2006 for a review of European
underground laboratories and Bettini 2007
for a recent compendium of underground laboratories worldwide).

The following years will lead to clearer picture of
a task distribution between European sites.
ApPEC will help finding solutions in case of 
conflicting national preferences and prioritizing possible extensions 
of the underground labs in accordance with the 
actual needs in Europe and worldwide.

\section{The high energy universe}

\begin{figure}[h]
\includegraphics[width=.45\textwidth]{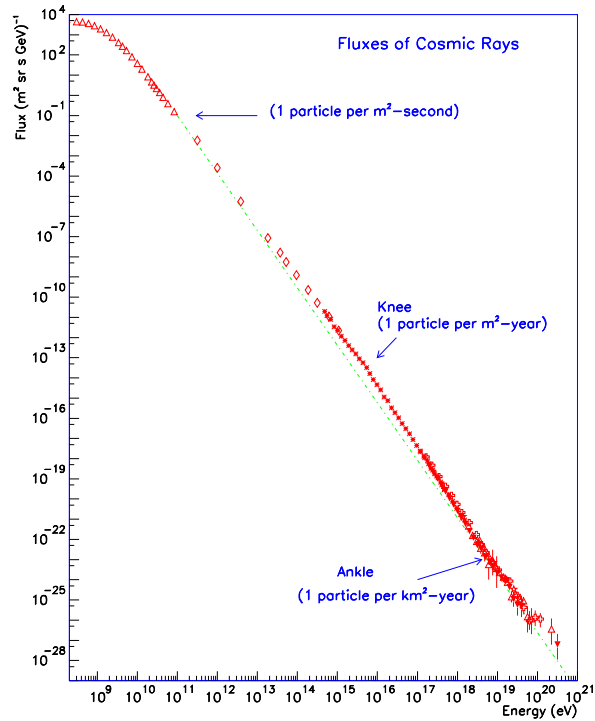}
\caption{
The spectrum of cosmic rays (courtesy S.P. Swordy). 
The region below 10$^{14}$ eV is the domain 
of balloon and satellite experiments, at higher energies ground based 
techniques take over.  Galactic supernova remnants can accelerate particles 
to energies of $10^{16}$-$10^{17}$\,eV, well above the "'knee"'. 
Galactic sources are believed 
to run out of power at $10^{17}$-$10^{18}$\,eV. 
Highest observed energies dwarf the 
Large Hadron Collider at CERN which will accelerate protons to $10^{13}$\,eV.
}
\label{CR}
\end{figure}

Much of classical astronomy and astrophysics 
deals with thermal radiation, emitted 
by hot or warm objects such as stars or dust. 
The hottest of these objects, such 
as hot spots on the surfaces of neutron stars, 
emit radiation in the range of some 
10$^3$ to 10$^4$ eV, about  thousand times more 
energetic than visible light. We know, 
however, that non-thermal phenomena, involving much higher energies, play an 
important role in the cosmos. First evidence for such phenomena came with the 
discovery of cosmic rays by Victor Hess in 1912. 
Hess measured radiation levels 
during balloon flights and found a significant increase with height, which he 
correctly attributed to a hitherto unknown penetrating radiation from space. 
In 1938, Pierre Auger proved the 
existence of extensive air showers -- cascades 
of elementary particles -- initiated by primary particles with energies above 
10$^{15}$ eV by simultaneously observing the 
arrival of secondary particles in Geiger 
counters many meters a part. Modern cosmic-ray detectors reveal a cosmic-ray 
energy spectrum extending to 10$^{20}$ eV and beyond 
(see Fig.\ref{CR}). That are 
breath-taking energies, a hundred million times above that of terrestrial 
accelerators (Watson 2006, Olinto 2007). 

How can cosmic accelerators boost 
particles to these energies? What is the nature 
of the particles?  Do the particles at the very highest energies
originate from the decay of super-heavy particle rather than
from acceleration processes (top-down versus bottom-up scenarios)?

The mystery of cosmic rays is going to be solved by an interplay 
of detectors for high energy gamma rays, charged cosmic rays and neutrinos.

\subsection{Charged cosmic rays} 

\vspace{1mm}
\noindent
{\it The highest energies}
\vspace{1mm}

\begin{figure}[h]
\center{
\includegraphics[width=.30\textwidth]{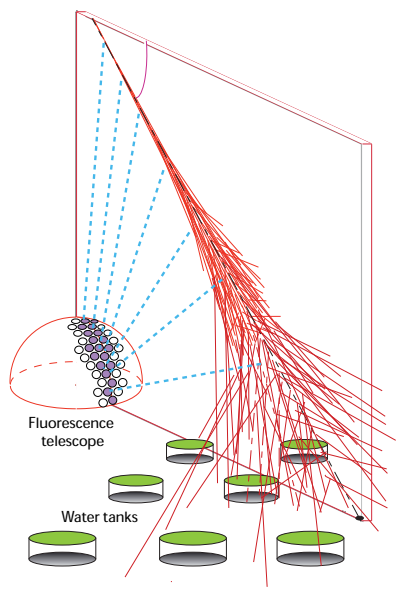}
}
\caption{
The Pierre-Auger detection principles: Fluorescence light  from air showers 
is recorded by telescopes,  particles at ground 
level are recorded by Cherenkov water tanks}
\label{Auger}
\end{figure}

The present flagship in the search for sources of 
ultra-high energy cosmic rays 
is the Southern Pierre Auger Observatory in Argentina 
(Abraham et al. 2004), a 3000-km$^2$ array 
of water tanks, flanked by air fluorescence telescopes, 
which measure direction 
and energy of giant air showers (see Fig.\ref{Auger}).

\begin{figure}[h]
\center{
\includegraphics[width=.45\textwidth]{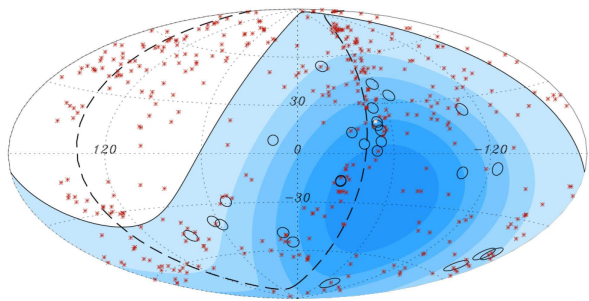}
}
\caption{
Sky map in galactic coordinates of 27 cosmic rays with highest energies 
detected by the 
Pierre Auger Observatory (black circles), compared to positions of 
472 quasars and active galactic nuclei with distance $<$ 75 Mpc. 
The dashed line marks the 
super-galactic plane. 
}
\label{Augermap}
\end{figure}

Even at energies above 10$^{19}$\,eV, where the cosmic flux is 
only about one particle per 
year and square kilometer, the Auger Observatory
can collect a reasonable number of events.  
Starting with these energies, the deflection of char\-ged particles in cosmic 
magnetic fields is going to be negligible and source tracing becomes possible. 
Very recently, the Auger collaboration has published a first sky map of events 
with energies above $10^{19.6}$ eV (Abraham et al. 2007, 
see Fig.\ref{Augermap}). There is a clear correlation of events 
with the super-galactic plane. 
Also, the authors report a correlation with positions 
of Active Galactic Nuclei with at least 99\% confidence level. Such a correlation 
would be in agreement with theoretical expectations which 
classify only two objects 
to be able to accelerate 
particles up to $10^{20}$ eV or higher: the jets of AGN and 
Gamma Ray Bursts. Whether the interpretation of AGN as sources of the observed 
cosmic rays will withstand further tests and higher statistics has to be seen. 
If confirmed,  
it would mark the first step into astronomy with charged cosmic rays.

Full-sky coverage would be obtained by a 
Northern observation site, which has been 
determined to be in Colorado/ USA. 
This array would significantly exceed the Southern 
detector in size, possibly at the expense of low energy sensitivity, i.e. using a 
larger spacing of detector elements which would result 
in a higher energy threshold. 
European groups will play a significant role to establish the scientific case, 
and after its full demonstration make a significant contribution to the design 
and construction of Auger-North. The cost is estimated at 90 Million
Euro, 
with a 45\% European contribution, start of 
construction is conceived for 2010. 

\vspace{3mm}
\noindent
{\it Between knee and ankle}
\vspace{1mm}

The subjects of the Pierre Auger Observatory 
are extragalactic sources. There are 
hardly sources in our own Galaxy which could 
accelerate particles ~above $10^{19}$ eV. 
Still, it remains interesting at which energy galactic 
sources actually run out of 
power. The energy range between 
$10^{16}$ and $10^{18}$ eV has been covered by very few 
experiments. 
Energy spectra determined by different experiments differ significantly, 
mostly due to the problems 
in proper energy calibration. On the other hand the 
region above $10^{16}$ eV is of crucial 
importance for our understanding of the origin 
and propagation of cosmic rays in the Galaxy.  What is the mass composition?  
Is this region dominated by sources other 
than supernova remnants?  Is there an 
early onset of an extragalactic component?  
What is the relation between cut-off 
effects due to leakage out of the 
Galaxy and cut-off effects due to maximum energies 
in sources? Three experiments (Kascade-Grande in Karlsruhe/Germany, Tunka-133 
in Siberia 
and IceCube/IceTop at the South Pole), each with about 1 km$^2$ area, 
are exploring this energy range. They will yield a precise 
measurement of the energy spectrum as well as improved knowledge about the 
mass composition (see e.g. Kampert 2006).

\vspace{3mm}
\noindent
{\it Below the knee}
\vspace{1mm}

At energies below the knee, one notes the recent successful launch of the 
Pamela satellite experiment. It will hopefully 
be followed in a few years by the launch of the much larger AMS spectrometer 
and its operation on the International Space Station ISS. 
The plans for AMS are strongly affected by the unclear 
situation for Space Shuttle missions.  Pamela and AMS, as well as future 
balloon missions, will search for anti-nuclei with much increased sensitivity 
and also measure the energy spectrum of different nuclei below the knee, up to 
energies  $10^{12}$-$10^{15}$ eV/nucleon (see for a science summary
Picozza \& Morselli 2006). A large satellite mission (Nucleon) extending
the direct particle measurements to close to the knee is planned in 
Russia, with some Italian participation.

\subsection{TeV gamma rays}

In contrast to charged cosmic rays, gamma rays propagate straight; 
compared to neutrinos, they are easy to detect. 
This has made them a powerful tracer of cosmic processes. 

Among all the different techniques 
developed so far for gamma detection, primarily two 
have succeeded in providing 
catalogues with reliable source detections and spectral 
measurements: 
{\it satellite detectors} and ground based 
{\it Imaging Atmospheric Cherenkov Telescopes} (IACTs). 

The first steps of cosmic particle acceleration are studied with satellite 
detectors for MeV energies like INTEGRAL, and for MeV-GeV energies, where 
the EGRET satellite has revealed more than 300 sources of radiation. 
In 2008, the GLAST detector will be launched and is expected to provide an 
even richer view of the universe at energies up to several $10^{13}$\,eV. 

Due to the small area of detectors on satellites, at energies above a few 
tens of GeV they run out of statistics. The higher energies are the domain 
of ground-based Cherenkov telescopes, covering the range above hundred GeV 
with extremely large sensitivities. They record the Cherenkov light from air 
showers originating from gamma ray interactions in the atmosphere. Large 
dishes focus the light to arrays of photomultipliers ("cameras"). 
From the shower image, direction, energy and character of the  primary 
particle (hadron versus gamma ray) can be derived. 

\begin{figure}[h]
\center{
\includegraphics[width=.45\textwidth]{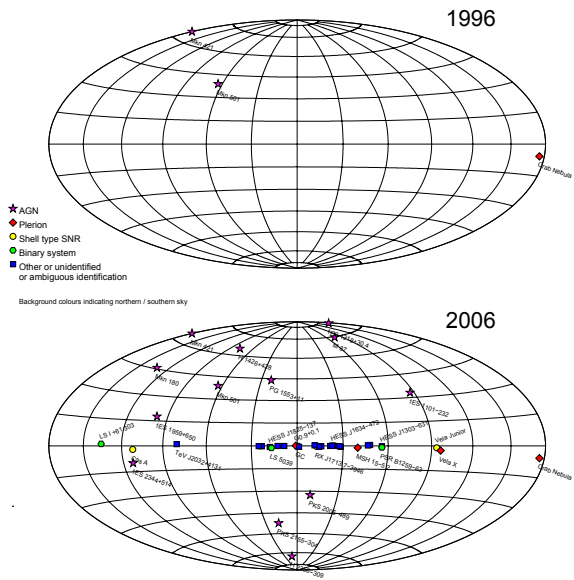}
}
\caption{
The TeV gamma sky in 1996 (top) and 2006 (bottom); 
courtesy K.\,Bernl\"ohr.
%Meanwhile, the number of 
%sources has grown to more than seventy. 
}
\label{gammasky}
\end{figure}

The IACT technique was pioneered in the USA with the development of the 
Whipple Telescope. Actually, it took the Whipple group nearly 20 years to 
eventually detect in 1989 a first source, the Crab Nebula (Weekes et al. 1989). 
During the last decade, European groups have been leading the development 
of IACTs and the field of ground-based high-energy gamma ray astronomy. 
Figure \ref{gammasky} 
shows a comparison of the TeV gamma sky map in 1996 and 2006. 
It illustrates the tremendous progress achieved within ten years. 
Most of the new sources in 
Fig. \ref{gammasky}  have been established by H.E.S.S., 
an array of four Cherenkov telescopes in Namibia, and MAGIC, a large 
telescope at La Palma (Voelk 2006, Aharonian2007). Both telescopes 
are being upgraded. 

\begin{figure}[h]
\center{
\includegraphics[width=.45\textwidth]{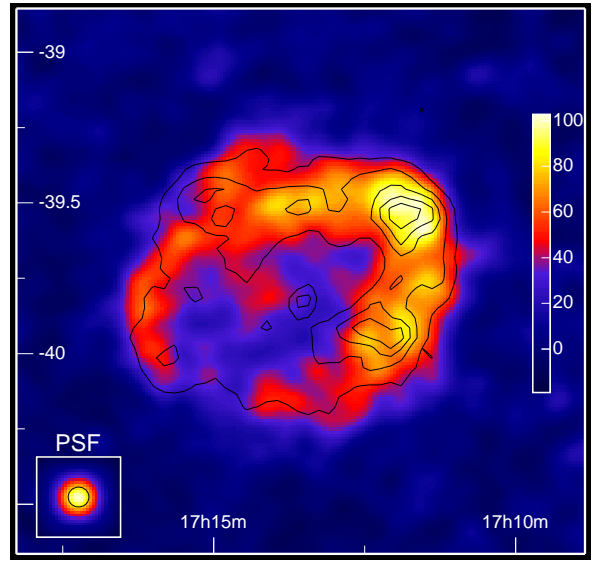}
}
\caption{
Supernova remnant RXJ1713.7-3946 at 
radio wavelength (contours) and TeV energies (coloured regions). 
}
\label{RX}
\end{figure}

IACTs have by now discovered more than seventy emitters of gamma rays at the 
$10^{11}$ to $10^{13}$ eV scale, many of them lining the Milky Way and 
revealing 
a complex morphology (see e.g.
Fig. \ref{RX}, taken from Aharonian et al. 2005).  
Most of the TeV sources correspond to 
known objects like binary stellar systems or supernova remnants. 
Others are still entirely unknown at any other wavelength and obviously 
emit most of their energy in the TeV range ("dark accelerators"). 
Going outside our own Galaxy, a large number of Active Galactic Nuclei 
have been observed and their fast variability demonstrated. 

While the results achieved with current 
instruments are already very impressive, 
they just give a taste of the TeV cosmos. 
The detailed understanding of the underlying processes and the 
chance to cover more than the "tip of the iceberg" can be improved 
dramatically by a much larger array of telescopes based on now well 
established techniques and observation strategies. A proto-collaboration 
has been formed to work jointly towards the design and realisation 
of such an instrument, which was christened CTA (Cherenkov Telescope Array,
see Hermann et al. 2007). 
It involves all European groups currently participating in IACTs, 
as well as a large number of additional new partners from particle physics and 
astrophysics.  Actually, CTA is on the list of emerging projects compiled 
by the {\it European Strategy Forum for Research Infrastructures (ESFRI)} 
and has been proposed by the ApPEC steering committee to be promoted 
to the status of a full ESFRI entry.

\begin{figure}[h]
%\center{
\includegraphics[width=.48\textwidth]{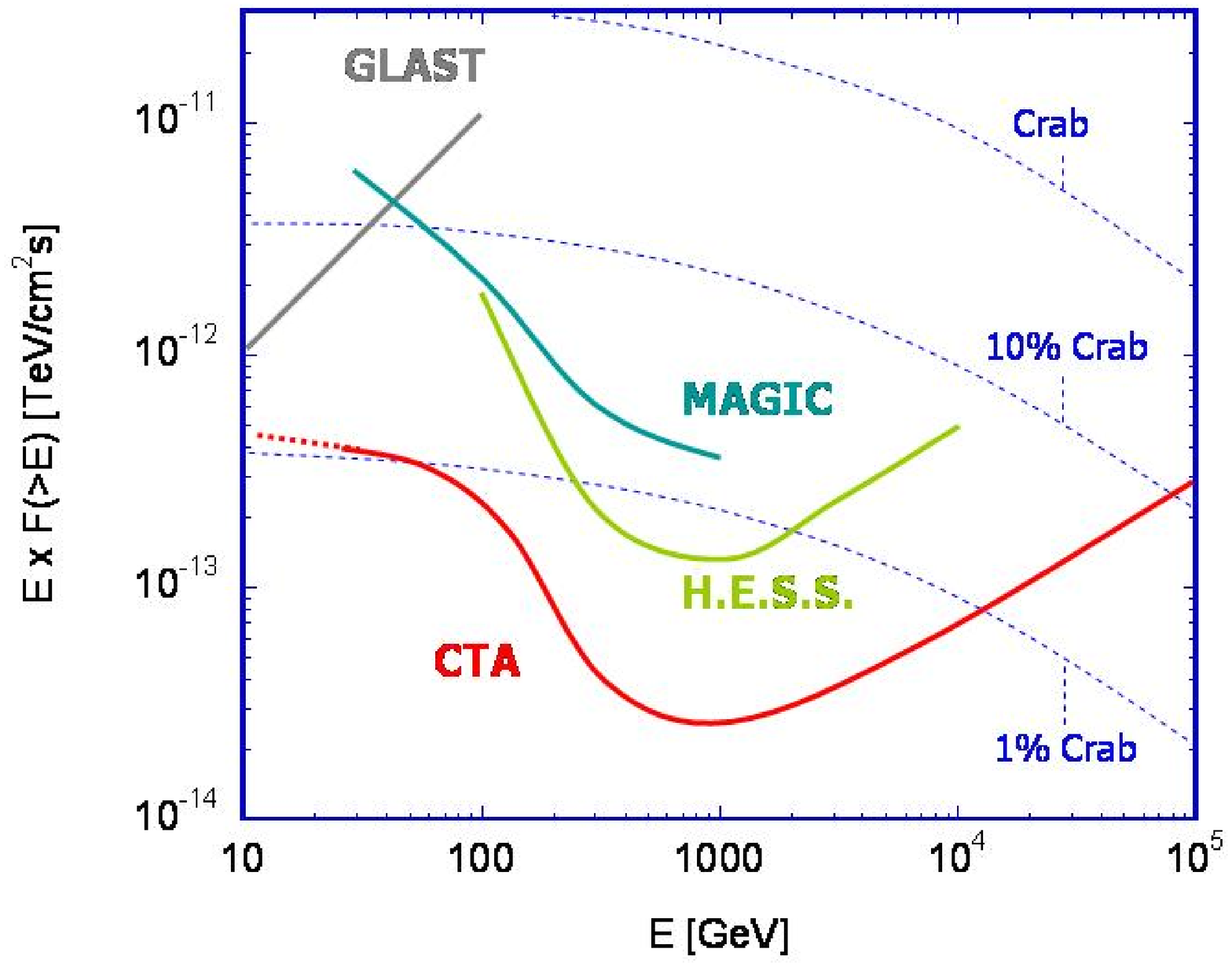}
%}
\caption{
Sensitivity of latest-generation Cherenkov instruments (H.E.S.S. and MAGIC) 
in comparison to that of the next-generation satellite experiment GLAST, 
and to the envisaged sensitivity of a next-generation Cherenkov instrument.
The final values for CTA will depend on the actual layout. 
For reference, the gamma ray flux from the Crab Nebula is shown.
}
\label{CTA}
\end{figure}

The goal of CTA is simultaneously 
increasing the energy bandwidth towards lower 
and higher energies, 
improving the sensitivity at currently accessible energies, 
and providing large statistics of highly constrained and very well 
reconstructed events (see Fig. \ref{CTA}, taken from Hermann et al. 2007). 
CTA will likely consist of a few 
very large central dishes providing superb efficiency below 50 GeV, 
embedded in an array of medium dishes giving high performance around a TeV, 
the latter being surrounded by a few-km$^2$ array of small dishes to catch the 
bright but rare showers at 100 TeV: altogether 40-70 telescopes. 
A similar concept (AGIS) is being discussed in the USA, and the need 
for cooperation and coordination is obvious.

CTA is conceived to cover both hemispheres, with one site in each. 
The field of view of the Southern site includes most of the Galaxy, 
the Northern telescope would instead focus to extragalactic objects. 
At energies above a few tens of TeV and over Mega-parsec distances, 
gamma rays are absorbed by the cosmic infrared light fields, and above 
a few hundreds of TeV by the 3K background. Therefore high energy 
sensitivity of the Northern ("extra-galactic") site is less important 
than for the Southern ("center of the galaxy") site. For the Southern site, 
emphasis would be put to high-energy sensitivity and excellent angular 
resolution in order to study the morphology of galactic objects.

CTA, with its many telescopes of small field of view, will likely ~be 
complemented by ~wide-angle devices like HAWC, the successor of the MILAGRO 
experiment in the USA (Sinnis, Smith \& McEnery 2004). This is a large
water pool with photomultipliers detecting the light from
air shower particles entering the water.
Compared to IACTs, HAWC would have a higher threshold and worse 
flux sensitivity but -- due to its large field of view -- better survey 
capabilities and better sensitivity to extended sources. From the low 
energy side, the US-initiated GLAST instrument would overlap with CTA. 
Since parallel observations are desirable and GLAST will be launched 
very soon, CTA construction should start as early as possible.

\subsection{High energy neutrinos} 

\begin{figure}[h]
\center{
\includegraphics[width=.48\textwidth]{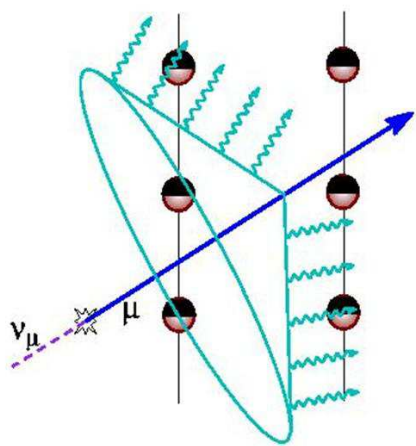}
}
\caption{
Neutrino telescopes consist of large arrays of photomultiplier 
tubes underwater or under ice. They detect the Cherenkov light emitted by 
charged particles which have been produced in neutrino 
interactions -- here from an 
up-going muon which stems from a neutrino having crossed the Earth.
}
\label{nutel}
\end{figure}

The physics case for high energy neutrino astronomy is obvious: 
neutrinos can provide an uncontroversial proof of the hadronic character 
of the source; moreover they can reach us from cosmic regions which are 
opaque to other types of radiation (Waxman 2007).  However, whereas 
neutrino astronomy in the energy domain of MeV has 
been established with the impressive observation of solar neutrinos 
and neutrinos from supernova SN-1987A, neutrinos with energies of 
Giga electron volts and above, which must accompany the production 
of high energy cosmic rays, still await discovery. Detectors underground 
have turned out to be too small to detect the corresponding feeble 
fluxes. The high energy frontier of TeV 
and PeV (1 PeV= $10^{15}$ eV) is currently being tackled 
by much larger, expandable arrays constructed in deep, open water or 
ice. They consist of photomultipliers detecting the Cherenkov 
light from charged particles produced by neutrino interactions 
(see Fig.\ref{nutel}). 
Flux estimations from astrophysical sources suggest that detectors on 
the cubic kilometre size scale are required for clear discoveries. 

European physicists have played a key role in construction and operation 
of the two pioneering large neutrino telescopes, NT200 in Lake Baikal
(Belolaptikov et al. 1997) 
and AMANDA at the South Pole
(Andres et al. 2001), and are also strongly involved in 
AMANDA's successor, IceCube (Ahrens et al. 2003, Spiering 2005).  

\begin{figure}[h]
\center{
\includegraphics[width=.45\textwidth]{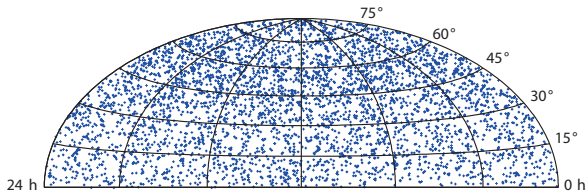}
}
\caption{
Sky map of 4282 events recorded by AMANDA in 2000-2004.
}
\label{Amasky}
\end{figure}

\begin{figure}[h]
\center{
\includegraphics[width=.40\textwidth]{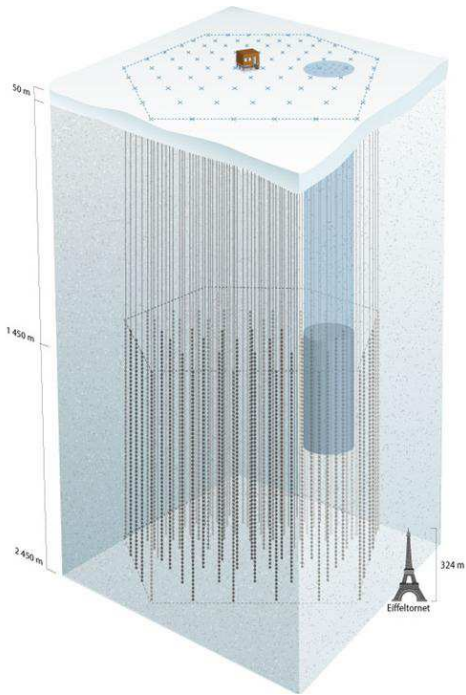}
}
\caption{
IceCube schematic view. 
IceCube consists of 80 strings each equipped with 60 photomultipliers 
between 1400 and 2400 meters depth. 
AMANDA (small cylinder) is integrated into IceCube. 
IceCube is complemented by a surface air shower array, IceTop, 
which records air showers and greatly enhances the physics 
capabilities of the deep ice detector.
}
\label{IceCube}
\end{figure}

Figure \ref{Amasky} shows a sky plot of the 4282 events recorded by AMANDA 
over five years (Achterberg et al. 2007).  Even with this highest statistics of 
high energy neutrino events ever collected, no point source signal 
could yet be identified, motivating the construction of detectors 
more than one order of magnitude beyond AMANDA size. Such a cubic 
kilometre detector, IceCube, is presently being deployed at the South Pole
(Ahrens et al. 2003). 
Completion is foreseen in January 2011; it then will consist of 4800 
photomultipliers arranged in 80 strings (see Fig.\ref{IceCube}) 
half of which have been installed by February 2008.

Using the Earth as a filter, IceCube observes the Northern sky. 
Complete sky coverage, in particular of the central parts of the 
Galaxy with many promising source candidates, requires a cubic kilometre 
detector in the Northern hemisphere (Halzen 2007). 
A prototype installation of AMANDA size, ~ANTARES, ~is presently being 
installed close to Toulon/France (Kouchner 2007), 
with 10 of a total of 12 strings 
already operating. R\&D work towards a cubic kilometer detector 
is also pursued at two other Mediterranean sites, ~the one (NEMO) 
close to Sicily, ~the other (NESTOR) close to the Peloponnesus
~(Spiering 2003, Katz 2006, Amore 2007). 

\begin{figure}[h]
\center{
\includegraphics[width=.37\textwidth]{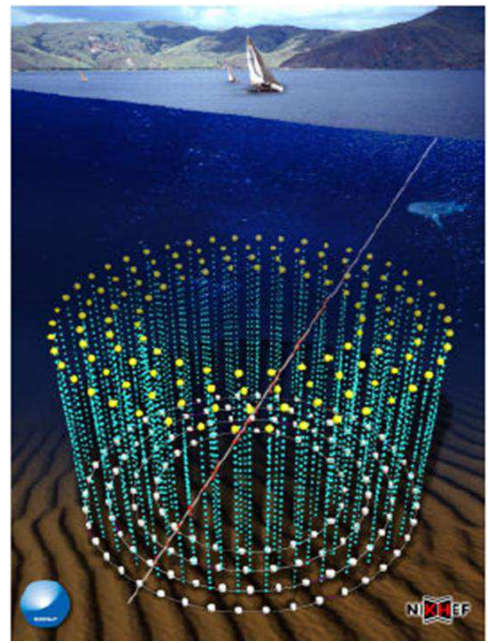}
}
\caption{
Artists view by M. Kraan (NIHKEF) of a possible design of KM3NeT.
}
\label{KM3}
\end{figure}

Resources for a cubic kilometre Mediterranean detector will be pooled 
in a single, optimized large research infrastructure. 
An EU-funded 3-year study (KM3NeT) is in progress to work out 
the technical design of a neutrino observatory in the Mediterranean, 
with construction envisaged to start in 2011 (Katz 2006). 
ESFRI has included KM3NeT 
in the {\it European Roadmap for Research Infrastructures}, 
thus assigning high priority to this project. 
Start of the construction of KM3NeT is going to be preceded by 
the successful operation of small scale or prototype detector(s) 
in the Mediterranean. Its design should also incorporate 
the improved knowledge on galactic sources as provided by  
gam\-ma ray observations, as well as initial results from IceCube --
including e.g. the possibility to construct, for similar cost,
a 3 or 5 times larger array with higher energy threshold. 
Still, the time lag between IceCube and KM3NeT should be 
kept as small as possible. Figure \ref{KM3} shows an example
for a possible KM3NeT configuration, based on a ``hollow
cylinder'' structure. 

\vspace{3mm}
\noindent
{\it Techniques for extremely high energies:} 
\vspace{1mm}

Emission of Cherenkov light in water or ice provides a relatively strong 
signal and hence a relatively low energy threshold for neutrino detection. 
However, the limited light transmission in water and ice requires a large 
number of light sensors to cover the required detection volume. 
Towards higher energies, novel detectors focus on other signatures of 
neutrino-induced charged particle cascades, which can be detected 
from a larger distance -- see B\"oser \& Nahnhauer 2006 for
a review. ~~Methods include recording the Cherenkov radio 
emission or acoustic signals from neutrino induced showers, as well as 
the use of air shower detectors responding to showers with a 
"neutrino signature". The very highest energies will be covered 
by balloon-borne detectors recording radio emission in terrestrial 
ice masses, by ground-based radio antennas sensitive to radio emission 
in the moon crust, or by satellite detectors searching for fluorescence 
light from neutrino-induced air showers. Taken all together, these 
detectors cover an energy range of more than twelve decades, starting 
at $10^{13}$-$10^{14}$\,eV (10-100 GeV) and extending beyond $10^{22}$ eV.  
Limits come i.e. from air shower detection with optical methods,
from radio searches for particle showers in the Moon, and from radio
searches for particle showers in ice. All of them have been derived within the
last decade. Exploitation of the full potential of these methods
needs large-scale R\&D work.

\vspace{3mm}
\noindent
{\it Summary on high energy neutrino detection}
\vspace{1mm}

Within the last five years, experimental sensitivities over the whole 
energy range 
have improved by more than an order of magnitude,
much faster than during the previous decades. 
%as shown in Fig. \ref{move}. 
Over the next 7-10 years, flux sensitivities 
are expected to move further down by a factor of 30-50, over the entire 
range from tens of TeV to hundreds of EeV.  
This opens up regions with high discovery potential.

\section{Gravitational Waves}

Gravitational waves would provide us with information on strong field 
gravity through the study of immediate environments of black holes. 
Typical examples are coalescences of binary systems of compact objects 
like neutron stars (NS) or black holes (BH). Even more spectacular events 
could be observed from galaxy collisions and the subsequent mergers of 
super-massive black holes residing in the centres of the galaxies. 
Further expected sources are compact objects spiralling into 
super-massive black holes, asymmetric supernovae, and rotating asymmetric 
neutron stars such as pulsars. Processes in the early Universe, on the 
time and length scales of inflation, must also produce gravitational waves.

Since the expected wavelengths are of the order of the source size, 
frequencies range from below a milli-Hertz to above a kilo-Hertz. 
Study of the full diversity of the gravitational wave sky therefore 
requires complementary approa\-ches: Earth-based detectors are typically 
sensitive to high-frequency waves, 
while space-borne detectors sample the low-frequency regime.

Pioneers of direct observations have been using resonant bar detectors, 
and some (significantly improved) bar detectors are still in operation. 
However, the most advanced tools for gravitational wave detection are 
interferometers with kilometre-long arms. The passage of a gravitational 
wave differently contracts space along the two directions of the arms and 
influences the light travel time (Hong, Rowan \& Sathyaprkash 2005). 

\begin{figure}[h]
\includegraphics[width=.45\textwidth]{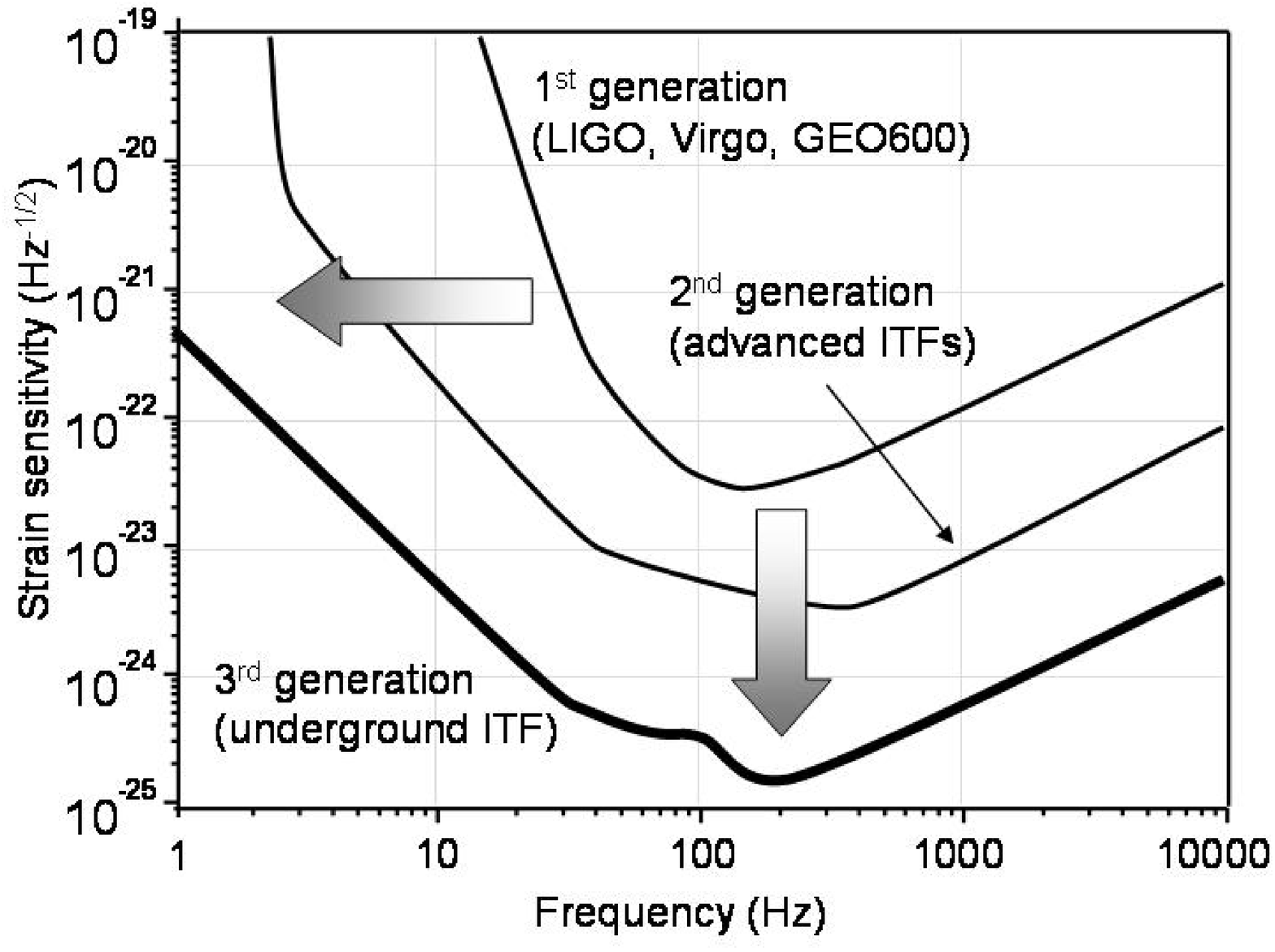}
\caption{
Current and expected sensitivities for ground-based 
gravitational waves detectors (Courtesy H. L\"uck).
Not shown are the curves for GEO-HF 
which will is the GEO600 interferometer tuned to high
frequencies and for DUAL, a 
future medium bandwidth detector called DUAL. The third generation 
interferometer curve is a very preliminary estimate. 
}
\label{gw1}
\end{figure}

At present, the world's most sensitive interferometer is LIGO (USA), 
the others being Virgo in Italy, 
GEO600 in Germany and the smaller TAMA in Japan.  
Given our current understanding of the expected event rates, gravitational 
wave detection is not very likely with these initial interferometers. 
Thus a mature plan exists for upgrades to the existing detectors 
systems to create {\it enhanced} 
and after that {\it advanced} detector systems, such 
that the observation of gravitational waves within the first weeks 
or months of operating the advanced detectors at their design 
sensitivity is expected (Fig. \ref{gw1}) 

\begin{figure}[h]
\center{
\includegraphics[width=.47\textwidth]{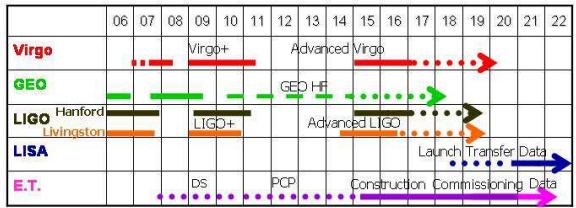}
}
\vspace{-2.4cm}
\caption{
Timeline of current detector operation and planned detector upgrades. 
The solid lines for the existing detectors indicate data taking times. 
In the regions of dotted lines the mode of operation is not yet defined. 
In the scenario shown, LISA would be launched in 2018. 
The {\it 3rd} generation plans start with a 3 year design study in 2008, 
followed by a 4 year preparatory construction phase. Construction and 
commissioning will last for 6 years and allow data taking from 2021 onwards.
}
\label{gw2}
\end{figure}

The European ground interferometers (GEO and Virgo) are turning to observation 
mode with a fraction of their time dedicated to their improvement 
(GEO-HF, Virgo+ and Advanced Virgo) --
see Fig. \ref{gw2}. Predicted event rates, e.g. for mergers 
of neutron star/black hole systems (BH-BH, NS-NS, NS-BH) are highly uncertain 
and range between 3 and 1000 for the "advanced" detectors planned to start 
data taking in about 5 years.

Even the advanced versions of the present interferometers will start reaching 
some fundamental limits, e.g. due to the seismic environment. 
Therefore the European Gravitational Wave Community envisages a {\it 3rd} 
generation interferometer as seismically quiet underground facility. 
The sensitivity target is an order of magnitude better than that of 
Advanced LIGO and Virgo (three orders of magnitude in event rate) 
with the seismic cut off going down to less than 1 Hz (see Fig. \ref{gw1}). 
This new facility (the "Einstein Telescope", {\it E.T.}) would be a dramatic 
step and allow Europe to play a key role in what will then be the field 
of gravitational-wave observational astronomy. 
E.T. would have a guaranteed rate of many thousand events per year 
and would move gravitational wave detectors in the category 
of astronomical observatories. 
A network of third and second generation detectors would measure to 
a few percent  the masses, sky positions and distances of binary black 
holes with stellar- and intermediate (i.e. a few hundred times solar) mass, 
out to a redshift of $z=2$ and $z=0.5$, respectively. 

E.T has been approved as a FP7 design study in 2007.  
The outcome of this work will be a conceptual design of the facility 
(including a selection of possible sites), followed by a more detailed 
preparatory construction phase to be in a position to start construction 
around 2015 (see Fig. \ref{gw2}). The design study will include conceptual 
aspects of the observatory to show that the envisaged sensitivity 
can be reached with the techniques, the funding and on the timescales 
foreseen. Cost for E.T. would be on the 500 Million Euro scale.

\vspace{3mm}
\noindent
{\it Gravitational wave astronomy from space}
\vspace{1mm}

\begin{figure}[h]
\center{
\includegraphics[width=.45\textwidth]{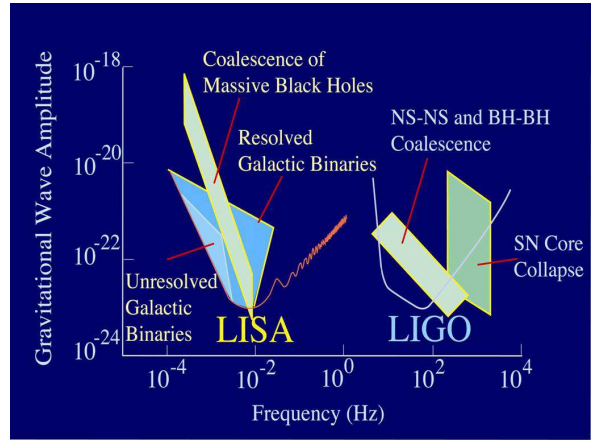}
}
\caption{
Comparison of the sensitivity regions defining the science potential
and the science targets
of earth-bound and space interferometers, respectively (Courtesy H. L\"uck).
}
\label{gw3}
\end{figure}

The frequency domain much below one Hz can be only explored from space. 
There is currently an ESA-NASA mission, LISA, which is scheduled for a 
launch in 2018. LISA would be ideally 
suited for the study of super-massive black holes mergers, galactic compact 
binaries (see Fig. \ref{gw3})
and potentially for the signatures of new physics beyond 
the standard model (Hughes 2007).

After transit to the final orbit, LISA
will be ready for data taking in 2020, roughly
coinciding with the Einstein Telescope, ET. LISA involves three 
spacecraft flying approximately 5 million kilometres apart in an 
equilateral triangle formation. These very long arms allow to cover 
a frequency range of $3 \cdot 10^{-5}$ to 1 Hz, complementary to the frequency 
window covered by ground-based instruments. Prior to LISA, the 
{\it LISA Pathfinder} mission, 
to be launched in 2010 by ESA, must test some of the critical 
new technology required for the instrument and proof the
feasibility of the concept.

The advanced ground based detectors will assure the detection of
gravitational waves within a few weeks or months. This is a necessary
condition to move ahead and to aquire the substantial funding
for E.T. and/or LISA. These two detectors then would move
gravitational wave instruments eventually into the league of true
astronomical observatories.

\section{The Big Picture}

In its strategy paper, the ApPEC roadmap committee argues that 
astroparticle physics is likely at the dawn of a golden age, as 
traditional astrophysics was two to three decades ago.  
The enormous discovery potential of the field stems from the fact 
that attainable sensitivities are improving with a speed exceeding 
that of the previous two decades. Improvement of sensitivities alone 
is arguably not enough to raise expectations. But on top of this, we 
are entering territories with a high discovery potential, as predicted 
by theoretical models. 

The increasing speed of progress is illustrated by
Fig. \ref{DM-Trend}, with a strongly
higher gradient of sensitivity improvement for Dark Matter search than any
time before. Curves of nearly identical shape can be drawn e.g.
for detection of high energy neutrinos or charged
cosmic rays. For the first time experimental and theoretical 
techniques allow -- or are going to allow -- forefront questions to be 
tackled with the necessary sensitivity. A long pioneering period during 
which methods and technologies have been prepared is expected to pay off 
over the next 5-15 years.  

The price tag of frontline astroparticle projects requires international 
collaboration, as does the realization of the infrastructure. 
~Cubic-kilometre neutrino telescopes, ~large gamma ray observatories, 
Megaton detectors for proton decay, or ultimate low-temperature devices 
to search for dark matter particles or neutrino-less double beta decay 
are in the range of 50-800 MEuro. 
Cooperation is the only way {\it a)} to achieve 
the critical scale for projects which require budgets and manpower 
not available to a single nation and {\it b)} to avoid duplication of 
resources and structures.  Astroparticle physics is therefore 
facing a similar concentration process as particle physics 
since several decades.

The following list of experiments on the $>$50 MEuro scale 
cost for investment is the
result of a first strategic approach of European
Astroparticle Physics (Phase-I of the ASPERA
Roadmap).

\begin{itemize}

\item {\bf High energy gamma astronomy:}\\
A large Cherenkov Telescope Array (CTA).
Desirably  two sites, one South, one North. Overall
cost 150-170 MEuro. Prototypes 2011, start of
construction in 2012.

 \item {\bf High energy neutrino astronomy:}\\
A kilometer scale neutrino telescope KM3NeT in the
Mediterranean, complementing IceCube on the opposite
hemisphere. 
Cost scale 200 MEuro. Start construction
after a preparatory phase, in 2011 or 2012.

\item {\bf High energy cosmic ray astronomy:}\\
~Auger-North, complementing the Pierre-Auger Site
in Argentina. Cost about 90 MEuro, 45\% of that from Europe.
Start construction in 2010 or 2011.

\item {\bf Direct Dark Matter searches:}\\
Two ``zero-background''dark matter experiments on the
ton scale, with a cost estimate of 150-180 MEuro for both
experiments and the related infrastructure together. Two different
nuclei and techniques (e.g. bolometric and noble liquid).
Decision in 2010/2011.

\item {\bf Masses and possible Majorana nature
of neutrinos by double beta decay experiments:}\\
Next generation experiments are ~GERDA, ~CUORE,
~Super-NEMO. Decision about an ``ultimate''
ton-scale experiment in the next decade, start of construction
not before 2013. Cost on the 150 MEuro scale.
Share with non-European countries.

\item {\bf Proton decay and low energy neutrino astrophysics:}\\ 
A detector on the Megaton scale. 
Worldwide collaboration, close coordination
with USA and Japan. Cost between 400 and 600 MEuro.
Decision on technology
after the end of the design phase, 2010-2012.

\item {\bf Gravitational waves:}\\
The third generation Graviational Wave interferometer,
located underground. Cost at the 500 MEuro scale. 
Need detection of Gravitational Waves with ``advanced'' interferometers
before construction would be approved.
Coordination with space plans (LISA) is important.

\end{itemize}
 
Naturally, there must be room for initiatives 
below the 50 Million Euro level. The Roadmap committee suggests that about 
20\% of astroparticle funding should be reserved for smaller 
initiatives, for participation in overseas experiments with non-European 
dominance, and for R\&D. Technological innovation has been a prerequisite 
of the enormous progress made over the last two decades and enabled maturity 
in most fields of astroparticle physics. It is also a prerequisite for future 
progress towards greater sensitivity and lower cost and must be supported 
with significant funds. 

With ApPEC and the related ERA-Net ASPERA, the process of coherent 
approaches within Europe has already successfully started. 
ApPEC represents nearly two thousand European scientists involved 
in the field. ApPEC helped to launch ILIAS, an Integrated 
Infrastructure Initiative with leading European infrastructures in 
Astroparticle physics. ILIAS covers experiments on double beta decay, 
dark matter searches and gravitational wave detection as well as 
theoretical astroparticle physics. ApPEC has also actively promoted 
the approval of KM3NeT as ESFRI project and FP6 design study, of CTA 
as emerging ESFRI project, of the Megaton neutrino 
detector study LAGUNA and the 
{\it 3rd} generation gravitational interferometer 
E.T. as FP7 design studies and also the FP7 support 
for the preparatory phase of KM3NeT.  
ApPEC will also play an important role in forming a coherent landscape of 
the necessary infrastructures, in particular of the underground laboratories.  

Phase-I of the roadmap (http://www.aspera-eu.org) describes physics
case and status of astropraticle physics and formulates recommendations
for each of the subfields. In a second phase ~(Phase-II), ~detailed information
on time schedule and cost have been collected from all experiments.
There is also a census on the present funding level collected from
the national funding agencies. Phase-III has just started.
During this phase, a precise calendar for milestones and decisions
will be prepared. Also, priorities will be formulated,
based on different funding scenarios. A Phase-III roadmap
paper will be released in autumn 2008. This work will
provide the necessary input for the decisions on the
large projects of the list above. Clearly, the required
resources exceed the present funding level. The roadmap
committee of ApPEC is, however, convinced that the prospects in this field
merit a substantially increased support.

\section*{Acknowledgements}

I thank my co-authors of the ApPEC  
Roadmap committee: F. Avignone, 
J.Bernabeu, L. Bezrukov,  P. Binetruy, H. Bl\"umer,  
K. Danzmann,  F. v. Feilitzsch,  E. Fernandez, W. Hofmann, 
J. Iliopoulos, U.Katz, P.Lipari, M. Martinez, A. Masiero, 
B. Mours, F. Ronga, A. Rubbia, S. Sarkar, G. Sigl, 
G. Smadja, N. Smith and A.Watson. 
I also acknowledge discussions with and support of 
members of the ApPEC Steering committee, 
in particular T. Bergh\"ofer, M. Bourquin and S. Katsanevas.

\end{document}